\documentclass[twocolumn,tighten,times]{aastex63}
\usepackage{graphicx}

\newcommand\aastex{AAS\TeX}

\newcommand{\msol}{\mbox{$M_\odot$}}

\newcommand {\mo}{{M}_\odot}

\newcommand{\HI}{H\,{\sc i} }
\newcommand{\HIk}{H\,{\sc i}}
\newcommand{\hi}{\ifmmode{\rm HI}\else{H\/{\sc i}}\fi} 
\newcommand {\kms}{\ifmmode{\rm km \, s^{-1}}\else{$\rm km \, s^{-1}$}\fi}

\shorttitle{\aastex\ Ultra-faint Dwarf Galaxy in the Local Group: KK153}
\shortauthors{Xu et al.}
\begin{document}
\title{FAST Discovery of A Gas-rich and Ultra-faint Dwarf Galaxy: KK153}

\correspondingauthor{Jin-Long Xu}
\email{xujl@bao.ac.cn}

\author{Jin-Long Xu}
\affiliation{National Astronomical Observatories, Chinese Academy of Sciences, Beijing 100101, People's Republic of China}
\affil{Guizhou Radio Astronomical Observatory, Guizhou University, Guiyang 550000, People's Republic of China}
%\affil{CAS Key Laboratory of FAST, National Astronomical Observatories, Chinese Academy of Sciences, Beijing 100101, People's Republic of China}

\author{Ming Zhu}
\affiliation{National Astronomical Observatories, Chinese Academy of Sciences, Beijing 100101, People's Republic of China}
\affil{Guizhou Radio Astronomical Observatory, Guizhou University, Guiyang 550000, People's Republic of China}
%\affil{CAS Key Laboratory of FAST, National Astronomical Observatories, Chinese Academy of Sciences, Beijing 100101, People's Republic of China}

\author{Nai-Ping Yu}
\affiliation{National Astronomical Observatories, Chinese Academy of Sciences, Beijing 100101, People's Republic of China}
\affil{Guizhou Radio Astronomical Observatory, Guizhou University, Guiyang 550000, People's Republic of China}
%\affil{CAS Key Laboratory of FAST, National Astronomical Observatories, Chinese Academy of Sciences, Beijing 100101, People's Republic of China}

\author{Chuan-Peng Zhang}
\affiliation{National Astronomical Observatories, Chinese Academy of Sciences, Beijing 100101, People's Republic of China}
\affil{Guizhou Radio Astronomical Observatory, Guizhou University, Guiyang 550000, People's Republic of China}
%\affil{CAS Key Laboratory of FAST, National Astronomical Observatories, Chinese Academy of Sciences, Beijing 100101, People's Republic of China}

\author{Xiao-Lan Liu}
\affiliation{National Astronomical Observatories, Chinese Academy of Sciences, Beijing 100101, People's Republic of China}
\affil{Guizhou Radio Astronomical Observatory, Guizhou University, Guiyang 550000, People's Republic of China}
%\affil{CAS Key Laboratory of FAST, National Astronomical Observatories, Chinese Academy of Sciences, Beijing 100101, People's Republic of China}

\author{Mei Ai}
\affiliation{National Astronomical Observatories, Chinese Academy of Sciences, Beijing 100101, People's Republic of China}
\affil{Guizhou Radio Astronomical Observatory, Guizhou University, Guiyang 550000, People's Republic of China}
%\affil{CAS Key Laboratory of FAST, National Astronomical Observatories, Chinese Academy of Sciences, Beijing 100101, People's Republic of China}

\author{Peng Jiang}
\affiliation{National Astronomical Observatories, Chinese Academy of Sciences, Beijing 100101, People's Republic of China}
\affil{Guizhou Radio Astronomical Observatory, Guizhou University, Guiyang 550000, People's Republic of China}
%\affil{CAS Key Laboratory of FAST, National Astronomical Observatories, Chinese Academy of Sciences, Beijing 100101, People's Republic of China}

\begin{abstract}
Based on a high-sensitivity \HI survey using the Five-hundred-meter Aperture Spherical radio Telescope (FAST), we identified an isolated \HI cloud with a system velocity of $\sim$127.0 \kms, which is associated with an optical galaxy KK153 in space. The \HI gas of KK153 shows a typical disk-galaxy structure. Using the Baryonic Tully-Fisher relation, we obtained that the distance to  KK153 is 2.0$_{-0.8}^{+1.7}$ Mpc. Adopting such distance, we derived a stellar mass of 4.1$_{-2.6}^{+10.0}\times10^{5}$ \msol and a neutral gas fraction of 0.63, implying that KK153 is a gas-rich ultra-faint dwarf (UFD) galaxy in the Local Group or its outskirts. KK153 shows a cool ($\sim$200 K) and warm ($\sim$7400 K) two-phase neutral medium. The $g-r$ color distribution of KK153 suggests that new stars are mostly forming in its inner disk. The dynamical mass of KK153 is 6.9$_{-3.0}^{+5.5}\times10^{7}$ \msol, which is about 60 times larger than its baryonic matter. Detection of such a low-mass and gas-rich halo poses a challenge to the theory of cosmic reionization.
\end{abstract}


\section{Introduction} \label{sec:intro}
The prevailing Lambda Cold Dark Matter ($\Lambda$CDM) cosmological model has successfully explained the Universe's large-scale structure. Still, it faces several challenges when applied to smaller scales \citep{Bullock2017}. One of the challenges is the disparity between the $\Lambda$CDM predictions and observations of dwarf galaxy numbers in the Local Group \citep{Moore1999,Klypin1999}. The detection of a considerable number of dwarf galaxies, in conjunction with the refinement of theoretical models, has led to a substantial reduction in the disparity \citep{Sales2022}. However, the number of ultra-faint dwarf (UFD) galaxies remains largely unconstrained \citep{Tollerud2008,Sales2022}. 

The UFD galaxies are generally defined as galaxies below a stellar mass of $M_{\ast}\sim10^{5}M_{\odot}$. They are the oldest, most dark matter-dominated, and most metal-poor \citep{Simon2019}. The following explanations for the missing UFD galaxies are available now. (i) a considerable number of the UFD galaxies have not been detected in observations due to the limitation of telescope detection sensitivity \citep{Kauffmann1993}; (ii) the abundance of the UFD galaxies may be suppressed to a greater extent than expected, because the smallest subhalos may not have effectively formed stars due to reionization \citep{Ricotti2005}, supernovae feedback \citep{Gallart2021}, ram pressure and tidal stripping \citep{Blitz2000}. Hence, the extensive search for UFD galaxies may provide insights into the aforementioned physical processes and further constrain the cosmological model.

The UFD galaxies previously searched were all gas-free in the Local group. Recently, only three gas-rich UFD galaxies have been detected: Leo T \citep{Irwin2007}, Leo P \citep{Giovanelli2013}, and Pisces A \citep{Tollerud2015}. Compared with optical spectroscopy to determine the redshift of the UFD galaxies, neutral hydrogen (\HIk) surveys are one of the most efficient methods for searching the UFD galaxies. The gas-rich UFD galaxies have the characteristics of weak gas emission and narrow linewidth. Thus, high sensitivity and high-velocity resolution are key in searching for the UFD galaxies in a new \HI survey. 

Using the Five-hundred-meter Aperture Spherical radio Telescope (FAST), we set out to carry out a FAST extragalactic \HI survey (FASHI) with a high sensitivity \citep{{Zhang2024}}. One of the scientific objectives of this survey is to search for dark and weak galaxies.  In this paper, based on the \HI survey using the FAST, we have discovered a gas-rich UFD galaxy in the Local Group or its outskirts.

\section{\HI survey and data processing}
For a new \HI survey using FAST \citep{Jiang2019,Jiang2020}, the 19-beam array receiver system in dual polarization mode is used as the front end. It formally works in the frequency range from 1050 MHz to 1450 MHz. For the backend, we choose the Spec(W) spectrometer that has 65,536 channels covering a bandwidth of 500 MHz for each polarization and beam, resulting in a frequency resolution of 7.629 kHz and corresponding to a velocity resolution of 1.6 \kms at z=0. The FAST \HI survey uses the drift scan mode. An interval between two adjacent parallel scans in Decl. is $\sim$1.14$^{\prime}$. Besides, we set an integration time of 1 second per spectrum. The system temperature was about 22 K. For intensity calibration to antenna temperature ($T_{\rm A}$),  a noise signal with an amplitude of 10 K was injected for 1 s every 64 s. The half-power beam width (HPBW) is $\sim$2.9$^{\prime}$ at 1.4 GHz for each beam. The pointing accuracy of the telescope was better than 10$^{\prime\prime}$. The detailed data reduction is similar to \citet{Xu2021}.  A gain $T_{\rm A}/\it S_{v}$ has been measured to be about 16 K Jy$^{-1}$. The measured relevant main beam gain $T_{\rm B}/\it S_{v}$ is about 21 K Jy$^{-1}$ at 1.4 GHz for each beam, where $T_{\rm B}$ is the brightness temperature. Finally, the mean noise RMS is $\sim$1.0 mJy beam$^{-1}$ with a velocity resolution of about 1.6 \kms in our observed image.

%The FAST \HI observations of FAST J0139+4328 in the new survey were carried out on  August 2, 2021 (region ID: DEC+433059-6).

\section{Results}
\label{sect:results}
To search for UFD galaxies in the Local Group, we mainly identify compact \HI gas clouds with system velocities below 250 \kms. From the detected clouds, we found an isolated cloud with a heliocentric system velocity of about 127.0 \kms, as shown in Figure \ref{fig:UFD_galaxy-HI}. The isolated clouds are typically thought to have no relatively massive galaxies within a radius of 100 kpc \citep{Taylor2017}. By comparing with the DESI optical images from the Legacy Surveys Sky Viewer websites\footnote {http://viewer.legacysurvey.org}, an optical counterpart was found at the center of the \HI  cloud. Based on the SIMBAD database and its coordinates, we found that the optical counterpart is a galaxy, named KK153 (LEDA 41920). Due to its extreme faintness, there is currently no key information about this galaxy including redshift. We are providing all the information about this galaxy for the first time and still use his optical name.

Figure~\textcolor{blue}{1a}  shows the \HI column-density map of KK153, overlaid on the DESI RGB image in color scale. The  \HI integrated velocity ranges from 115.0 \kms to 142.0 \kms. The optical disk of KK153 displays an extended structure from south to north. We measured the galaxy's position angle (PA) to be approximately 167.0 degrees and the axial ratio ($q$=b/a) to be about 0.75. The inclination angle (${\it i}$) can be estimated by
$\rm sin^\mathrm{2} \it i =  (\rm 1- \it q^{\rm 2})/(\rm 1- \it q_{\rm 0}^{\rm 2})$, where the value of $q_{\rm 0}$ could depend on galaxy morphology. For irregulars, $q_{\rm 0}$ can be adopted as  0.4 \citep{Hunter2006}. We obtained that $i$ value of KK153 is 68.1$_{-6.2}^{+6.2}$ degrees. In Fig.~\textcolor{blue}{1b}, the zoom-in DESI-RGB image of KK153 shows that several stars can be distinguished in its optical disk. Furthermore, the \HI emission shows that KK153 has a compact gas structure. The $q$ value of the gas structure is measured to about 0.76, roughly equal to its optical value. The effective radius can be calculated with $R_{\rm eff} = \sqrt{S^{2}-B_{\rm F}^{2}}$/2, where  $B_{\rm F}$  is the beam size, and $S$ is the uncorrected \HI sizes of galaxies.  $S$ is adopted as the outmost size at 3$\sigma$ level for KK153 from Fig. \textcolor{blue}{1a}. We obtained that the ratio of gas to optical effective radius for KK153 is about 8.9, implying that its optical disc is deeply buried in the \HI gas. Moreover, we obtained that the outermost column-density value of KK153 is 4.2$\times$10$^{17}$ cm$^{-2}$, and its peak value is 1.1$\times$10$^{19}$ cm$^{-2}$.

\begin{table}
\centering
\caption{\small Measured and derived properties of KK153.}
\vspace{-8pt}
\label{tab:prop}
\setlength{\tabcolsep}{27pt}
\begin{tabular}{lcccc}
\noalign{\vspace{5pt}}\hline\hline\noalign{\vspace{5pt}}
Parameter & Value \\
\noalign{\vspace{5pt}}\hline\noalign{\vspace{5pt}}
R.A. & 12$^{\rm h}$35$^{\rm m}$01.7$^{\rm s}$  \\
Decl. & 58$^{\rm \circ}$23$^{\rm \prime}$11.8$^{\rm \prime\prime}$ \\
$V^{\rm W}_{\rm sys}$ ($\kms$) & 127.5$_{-0.2}^{+0.2}$ \\
$W^{\rm W}_{\rm 50}$ ($\kms$) & 18.4$_{-0.8}^{+0.8}$  \\
$W^{\rm W}_{\rm 20}$ ($\kms$) & 23.3$_{-0.9}^{+0.9}$  \\
$I^{\rm W}_{\rm tot}$ ($\rm mJy~\kms$) & 515.2$_{-49.5}^{+49.5}$ \\
$V^{\rm N}_{\rm sys}$ ($\kms$) & 127.8$_{-0.1}^{+0.1}$ \\
$W^{\rm N}_{\rm 50}$ ($\kms$) & 3.2$_{-0.6}^{+0.6}$  \\
$W^{\rm N}_{\rm 20}$ ($\kms$) & 4.0$_{-0.7}^{+0.7}$  \\
$I^{\rm N}_{\rm tot}$ ($\rm mJy~\kms$) & 38.2$_{-8.1}^{+8.1}$ \\
$i$ (deg) & 68.1$_{-6.2}^{+6.2}$  \\
$PA$ (deg) & 167.0$_{-1.7}^{+1.7}$  \\
$V_{\rm rot}$ ($\kms$) & 12.6$_{-0.5}^{+0.5}$  \\
%$\sigma_{v}$ ($\kms$) & 7.8$\pm$1.4  \\
$m_{\rm g}$ (mag) & 17.3$_{-0.1}^{+0.1}$  \\
$m_{\rm r}$ (mag) & 16.9$_{-0.1}^{+0.1}$  \\
$D$ (Mpc) & 2.0$_{-0.8}^{+1.7}$ \\
$R_{\rm opt}$ (pc) & 209.3$_{-83.7}^{+177.9}$  \\
$R_{\rm HI}$ (kpc) & 1.9$_{-0.8}^{+1.5}$  \\
$M_*$ (\msol) & 4.1$_{-2.6}^{+10.0}\times10^{5}$ \\
$M_{\rm \hi}$ (\msol) & 5.2$_{-3.3}^{+12.6}\times10^{5}$  \\
$M_{\rm dyn}$ (\msol) & 6.9$_{-3.0}^{+5.5}\times10^{7}$  \\
\noalign{\vspace{5pt}}\hline\hline\noalign{\vspace{5pt}}
\end{tabular}
Note: Equatorial coordinates (R.A., Decl.); system velocity ($V^{\rm N}_{\rm sys}$ and $V^{\rm M}_{\rm sys}$); line widths at 50\% of the peak flux ($W^{\rm W}_{\rm 50}$ and $W^{\rm N}_{\rm 50}$), and at 20\% of the peak flux ($W^{\rm W}_{\rm 20}$ and $W^{\rm N}_{\rm 20}$);  integral flux ($I^{\rm N}_{\rm tot}$ and $I^{\rm M}_{\rm tot}$); inclination angle ($i$); position angle ($PA$); rotation velocity ($V_{\rm rot}$); $g$-band and $r$-band  apparent magnitudes ($m_{\rm g}$, $m_{\rm r}$), which are corrected for Galactic extinction;  distance ($D$); optical effective radius ($R_{\rm opt}$); \HI effective radius ($R_{\rm eff}$);  stellar mass ($M_*$); \HI gas mass ($M_{\rm \hi}$); dynamic mass ($M_{\rm dyn}$).
\end{table}

\begin{figure*}[ht!]
\centering
\includegraphics[width=0.9\textwidth]{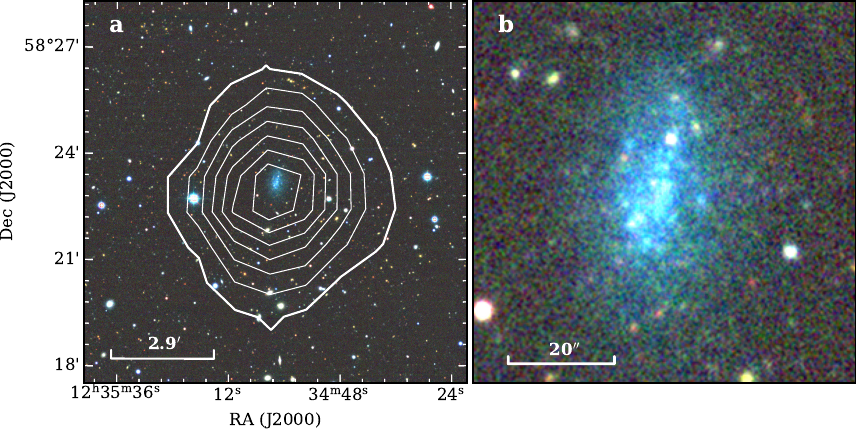}
\setlength{\abovecaptionskip}{0pt}
\caption{{\bf a}, \HI column-density map of KK153 from the FAST observation shown in white contours overlaid on the DESI-RGB ($g, r, z$) image in color scale. The white contours begin at 4.2$\times$10$^{17}$ cm$^{-2}$ (3$\sigma$) in steps of 2.0$\times$10$^{18}$ cm$^{-2}$.  {\bf b},  the zoom-in DESI-RGB image of KK153.}
\label{fig:UFD_galaxy-HI}
\end{figure*}

\begin{figure}
\vspace{6pt}
\centering
\includegraphics[width=0.45\textwidth]{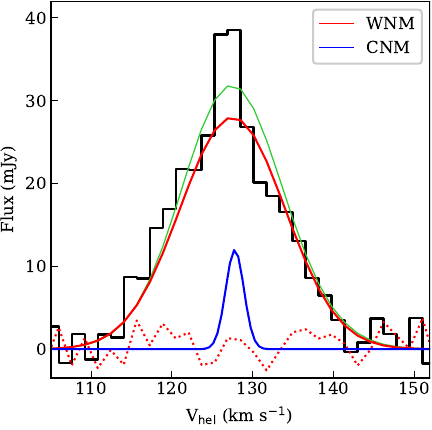}
\setlength{\abovecaptionskip}{0pt}
\caption{Global \HI profile of KK153 shown in a black line. The green line indicates the BusyFit fitting result, while the red and blue lines represent the double Gaussian fitting results. The residual after the Gaussian fitting is represented by a red dashed line.}
\label{fig:UFD_galaxy-HI}
\end{figure}

\begin{figure*}
\centering
\includegraphics[width=0.95\textwidth]{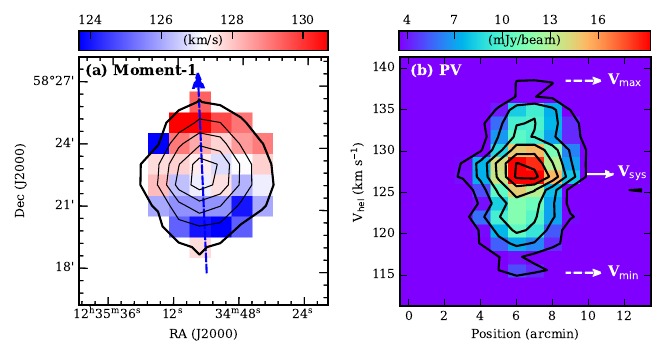}
\setlength{\abovecaptionskip}{-5pt}
\caption{a, velocity-field map of KK153 in color scale overlaid with its \HI column-density map in black contours. The black contours begin at 4.2$\times$10$^{17}$ cm$^{-2}$ in steps of 2.0$\times$10$^{18}$ cm$^{-2}$.  b and c, position-velocity (PV) diagrams of the observed data in color scale overlaid with the black contours. The black contours begin at 3$\sigma$ (3.0 mJy beam$^{-1}$) in steps of 3$\sigma$.  The cutting direction of the PV diagrams is indicated by the blue arrow in panel a.}
\label{fig:Dark_galaxy-HI}
\end{figure*}

Figure \textcolor{blue}{2} displays the global \HI profile of KK153. The \HI profile shows a rare shape. At the center position, there is a structure similar to the chimney. The identification of this special shape benefits from FAST's ability to have both high-velocity resolution and high sensitivity. The \HI profile can not be well fitted by a new analytic function BusyFit  \citep{Westmeier2014}, as shown in the green line in Fig. \textcolor{blue}{1b}. We fitted the \HI profile using double Gaussian and obtained a narrow Gaussian and a wide Gaussian components. The two-component Gaussian fit is strongly preferred to a BusyFit component. The fitting parameters are listed in Table \ref{tab:prop}. The velocity dispersion $\sigma = W_{50}/\sqrt{8ln(2)}$, where $W_{50}$ is the line width at 50\% of the peak flux. We obtained that $\sigma$ of the narrow Gaussian component 1.4$\pm$0.3 \kms,  while 7.8$\pm$0.4 \kms for a wide Gaussian component, indicating that KK153 has a cold ($\sim$200 K) and warm ($\sim$7400 K) two-phase neutral medium.  The temperature of a cool neutral medium (CNM) component is generally $\leq$1000 K in the \HI gas, while $\geq$5000 K for a warm neutral medium (WNM) component \citep{Young1996,Warren2012,Adams2018}. 

Figure \textcolor{blue}{3a} shows the velocity-field map (Moment-1) of KK153. This map shows a consistent velocity gradient across KK153 from south to north, as indicated by a blue arrow. Such a velocity gradient implies a rotating gas disk in KK153, but its main axis direction has an angle difference of about 17$^{\circ}$ with that of its optical disk.  To further understand the distribution of gas in the \HI gas disk, we made a position-velocity (PV) diagram across KK153 with a cutting direction along the main axis of its \HI disk, as shown in Fig. \textcolor{blue}{3b}. Through the PV diagram, we found that the gas disk of KK153 indeed shows a rotation structure. The measured minimum velocity ($V_{\rm min}$) is 115.0 \kms, and the maximum velocity ($V_{\rm max}$) is 139.1 \kms for its \HI rotation disk.

The Baryonic Tully-Fisher (BTF) relation links a galaxy's rotation velocity to its total baryonic mass \citep{McGaugh2012,McGaugh2015,Lelli2016}, which has been proven effective in determining the distance of galaxies, such as UFD galaxy Leo P \citep{Giovanelli2013}. Since KK153 shows a rotational gas disk, we can apply the BTF relation to estimate its distance. The rotation velocity can be calculated by $V_{\rm rot}=W_{20}/2\rm sin(\it i)$, where $W_{20}$ is linewidth at 20\% of the peak flux \citep{McGaugh2012}. Adopting the linewidth of WNM, we obtained $V_{\rm rot}$  of 12.6$\pm$0.5 \kms for KK153. To verify this value, the $V_{\rm rot}$ is also estimated by ($V_{\rm max}-V_{\rm sys})/\rm sin(\it i)$, as shown in Fig. \textcolor{blue}{3b}. We derived that $V_{\rm rot}$ is 12.5$\pm$0.8 \kms, which is nearly equal to that obtained from a linewidth. This indicates that the calculated $V_{\rm rot}$ is reliable. Here we use $V_{\rm rot}$ obtained from the linewidth to limit the distance of KK153. 

\begin{figure}
\vspace{0pt}
\centering
\includegraphics[width=0.48\textwidth]{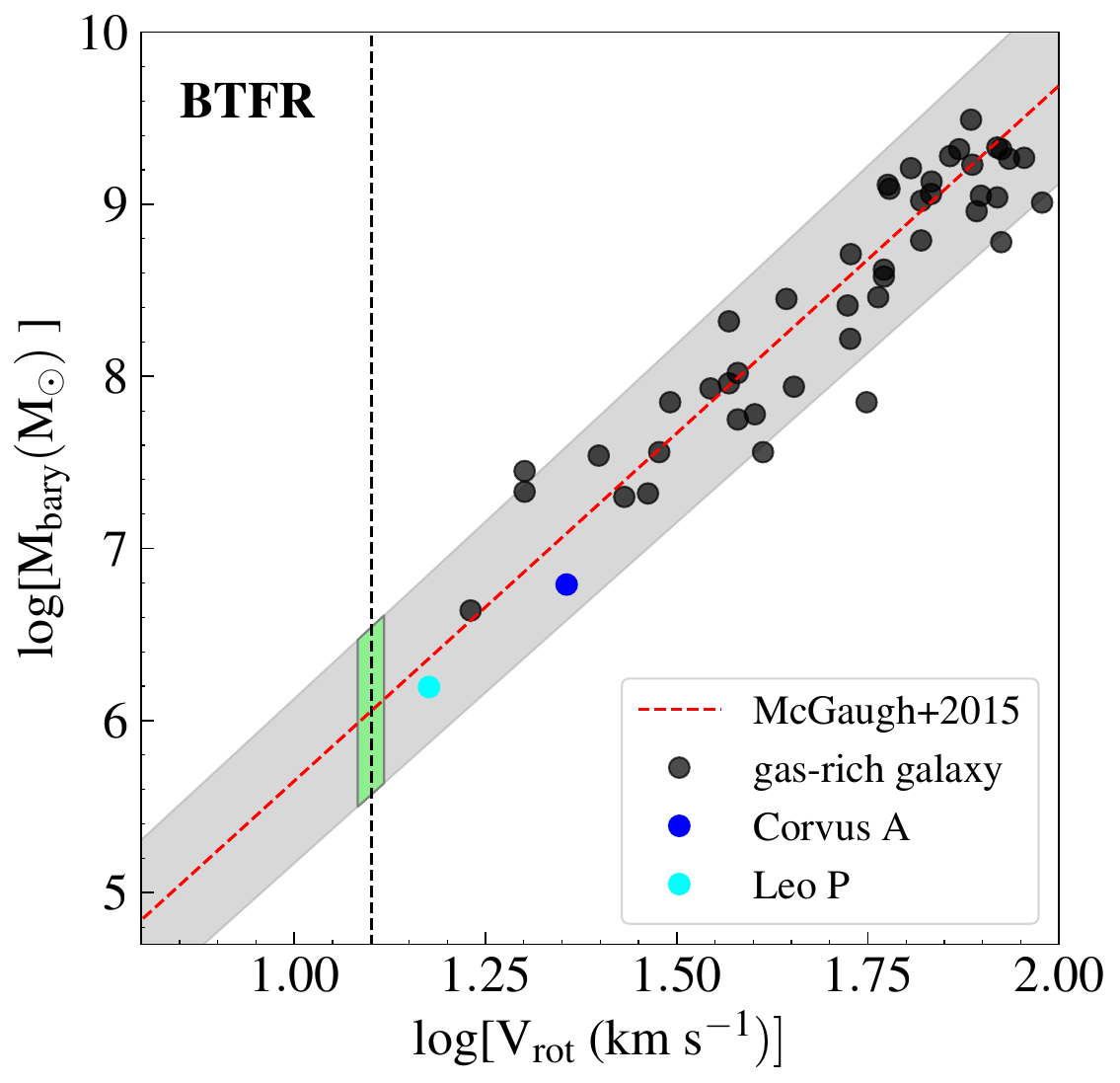}
\caption{(a), Baryonic Tully-Fisher relation (BTFR). The red dashed line represents a best-fit relation $\rm log(\it M_{\rm bar})=\rm a+b\times log(\it V_{\rm rot})$, where a=1.61$\pm$0.18 and b =4.04$\pm$0.09. The gas-dominated galaxies are marked by black points \citep{McGaugh2015}. The black dashed lines mark the rotation velocity of  KK153, while the green shaded area represents the range of its baryonic mass.}
\label{fig:Tully–Fisher}
\end{figure}

Figure \ref{fig:Tully–Fisher} illustrates the position of KK153 on the BTF relation based on its $V_{\rm rot}$. The green shaded area represents the range of baryonic mass for KK153. According to the BTF relation, we obtained the baryonic mass ($M_{\rm bary}$) of 1.1$_{-0.8}^{+2.8}\times10^{6}$ $\mo$ for KK153. Moreover, $M_{\rm bary}=M_*+M_{\rm gas}$. The total gas mass $M_{\rm gas} = 1.33\times M_{\rm HI}$, where the \HI gas mass $M_\mathrm{\hi}=2.36\times10^{5}D^{2}I_{\rm tot}$.  $D$ is the adopted distance in Mpc and  $I_{\rm tot}$ is the total integrated flux in Jy km s$^{-1}$, which contains WNM and CNM. The stellar masses ($M_*$) are estimated using the mass-to-light ratio equation  $\log(M_*/L_g)$=-0.601+1.294($g-r$), where $L_g$ is the $g$ band luminosity derived from the absolute magnitude \citep{Herrmann2016}. By using the above multiple expressions, we can deduce that the distance $D$ is 2.0$_{-0.8}^{+1.7}$ Mpc for KK153. To verify the effectiveness of the distance estimation method via the BTF relation, we also check a low-mass, isolated galaxy Corvus A at 3.5 Mpc \citep{Jones2024}. The distance to Corvus A is determined via the tip of the red giant branch standard candle. From Fig. \ref{fig:Tully–Fisher}, we see that Corvus A well follows the BTF relation. As a comparison, the distance to KK153 can be estimated to be 1.9$\pm$0.2 Mpc using the Cosmicflows-3 method \citep{{Kourkchi2020}} and the system velocity of about 127 \kms, which is consistent with that obtained from the BTF relation, placing KK153 within the Local Group or its outskirts.  

\section{Discussion and Conclusion}
\label{sect:discussion}

\begin{figure*}
\vspace{5pt}
\centering
\includegraphics[width=.46\textwidth]{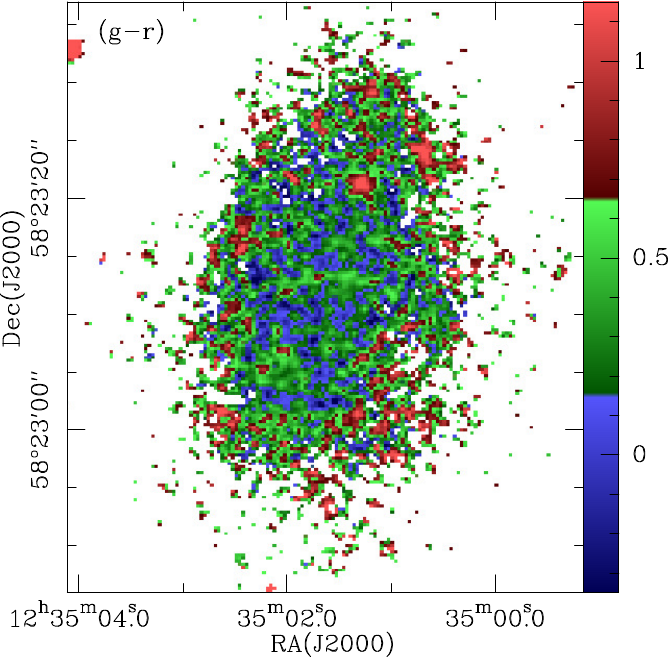} \hspace{0.8cm}
\includegraphics[width=.4\textwidth]{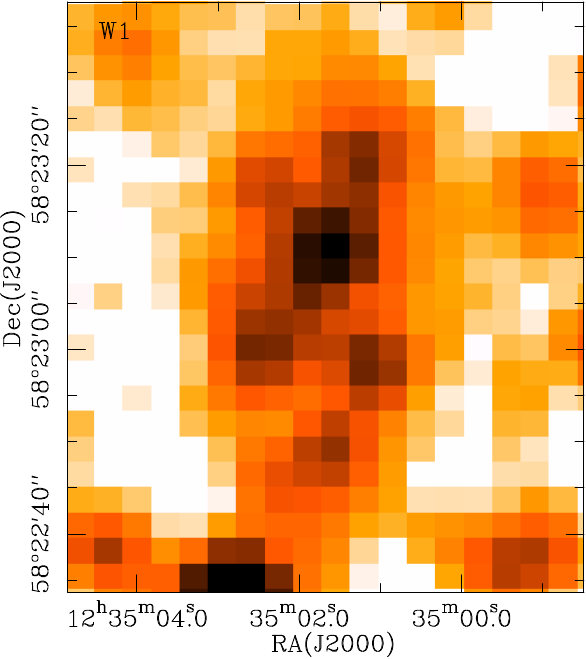} 
\caption{Left panel: the distribution of DESI \textit{g}-\textit{r} color in each pixel for KK153. Right panel: the WISE W1-band emission of KK153.}
\label{fig:g-r_bands}
\end{figure*}
Previously, three gas-rich UFD galaxies (Leo T, Leo P, Pisces A) were detected in the Local Group. For KK153, we obtained a distance of 2.0$_{-0.8}^{+1.7}$ Mpc from the BTF relation. Using the distance, we derived  $M_{\rm HI}$ of 5.2$_{-3.3}^{+12.6}\times10^{5}$ \msol and $M_*$ of 4.1$_{-2.6}^{+10.0}\times10^{5}$ \msol. The corresponding $M_{\rm HI}/M_*$ is about 1.3, which coincides with a typical value in low-mass dwarfs \citep{Lelli2022}. Moreover, a neutral gas fraction $f_{\rm gas} = M_{\rm gas}/(M_{\rm gas}+M_*)$, then we obtained $f_{\rm gas}$ of 0.63. Both $M_{\rm HI}/M_*$ and $f_{\rm gas}$ are independent of distance.  From the obtained $M_*$, $f_{\rm gas}$, and $M_{\rm HI}/M_*$, it suggests that KK153 is a gas-rich UFD galaxy. This will add a new gas-rich UFD galaxy in the Local Group or its outskirts. In the previous three gas-rich UFD galaxies, Leo T has the smallest stellar mass \citep{Irwin2007}. The stellar mass of KK153 is similar to that of Leo T. While KK153 is currently the only gas-rich UFD galaxy with obvious rotational characteristics, which makes it possibly the smallest gas-rich disk galaxy discovered so far.

The \textit{g}-\textit{r} color is generally used to determine a division of red to blue galaxies. Here we simply take \textit{g}-\textit{r} = 0.65 as a division from \citet{Nelson2018}. We obtained the comprehensive \textit{g}-\textit{r} color of 0.38 for  KK153 from Table \ref{tab:prop}, indicating that it is a blue galaxy that has recently formed stars. In Fig. \ref{fig:g-r_bands}, the left panel shows the distribution of  \textit{g}-\textit{r} color in each pixel for KK153. The value of the \textit{g}-\textit{r} color ranges from -0.57 to 1.53 with a median value of 0.41, implying that KK153 not only contains a large number of blue-color stars but also red-color stars. The median value is associated with the comprehensive \textit{g}-\textit{r} color of KK153. We found that the distribution of these color values is not disordered in KK153. The red-color stars are located on the periphery of the optical disk, while the blue-color stars are more likely to be distributed inside the disk.  In Fig. \ref{fig:g-r_bands}, the right panel displays the WISE W1-band (3.4 $\mu$m) emission of KK153. The WISE W1-band data is obtained from the NEOWISE program \citep{Mainzer2014}. The W1 band traces the continuum emission of the warm dust from evolved stars with minimum extinction. We see that several of the brightest evolved stars are located inside the disk of KK153. Compared with the left panel of Fig. \ref{fig:g-r_bands}, most of these evolved stars have a blue color. It indicates that the stars in the gas-rich UFD KK153 may form from the outside to the inside (``outside-in'' mode), which is associated with other low-mass ($M_*<10^{10}$ \msol) galaxies \citep{Perez2013,Pan2015}.  Previous studies found some massive star-forming galaxies with an outside-in assembly mode. They suggested that these galaxies are likely in the transitional phase from normal galaxies to quiescent galaxies \citep{Wang2017}. KK153 is in an isolated environment, implying that it is very likely also in the transitional phase. 

KK153 shows CNM ($\sim$200 K) and WNM ($\sim$7400 K) two-phase neutral medium. We obtained that the masses of the CNM and WNM in KK153 are 3.6$_{-2.3}^{+8.6}\times10^{4}$ \msol and 4.9$_{-3.1}^{+12.0}\times10^{5}$ \msol, respectively. The formation of the CNM in Leo T is due to the tidal effects of the Milky Way \citep{Adams2018}, while the CNM formation in the isolated galaxy KK153 may be due to the natural cooling of the WNM. This two-phased medium is typical of gas-rich faint dIrr galaxies \citep{Begum2006}. Some galaxies in the Local Group have a similar interstellar medium with two phases, like Leo A \citep{Young1996}, Sagittarius DIG \citep{Young1997}, and Leo T \citep{Adams2018}. However, the dwarf galaxy LGS 3, which has stopped making stars, does not have a CNM phase \citep{Young1997}. Molecular hydrogen ($\rm H_2$) is the main raw material for star formation in galaxies. We do not know if CNM is directly involved in star formation. Still, a certain amount of CNM has been detected in low-mass dwarf galaxies with forming stars, indicating that CNM may play an important role in galaxy formation. 

The deeper observations from the FAST trace an \HI disk to a larger radial extent, allowing us to probe more of the dark matter halo of KK153. From the virial theorem, the dynamical mass within the \HI extent can be estimated with $M_\mathrm{dyn} = V^{2}_{\rm rot}R_{\rm HI}/G$, where $V_{\rm rot}$ is rotation velocity, $R_{\rm HI}$ is effective radius, and $G$ is gravitational constant. We obtained that $M_{\rm dyn}$ of KK153 is 6.9$_{-3.0}^{+5.5}\times10^{7}$, which is about 60 times larger than its baryonic matter. Although dark matter dominates in KK153, its small mass gives it a shallow potential well and makes it susceptible to damage from external environments and internal supernovae. The intact shape of the gas disk of KK153 on the outside and the presence of a large amount of CNM inside indicate that this UFD galaxy has not been affected by the two dynamic processes mentioned above.  The reionization of the cosmic is achieved by evaporating the gas in the dark halo and suppressing the cooling of \HI gases so that dark halos with a mass less than $10^{9}$ \msol remain dark \citep{Quinn1996,Okamoto2008,Benitez2017,Herzog2023}. The dynamical mass of KK153 is at least one order of magnitude lower than the threshold value, but it has a certain amount of optical emission and is not very dark.  In addition, the diffuse gas with a column density below 5$\times10^{19}$ cm$^{-2}$ are susceptible to the cosmic background UV photons, making them difficult to survive \citep{Neff2005,Davies2006}. The obtained column density for KK153 range from 4.2$\times$10$^{17}$ cm$^{-2}$ to 1.1$\times$10$^{19}$ cm$^{-2}$. Due to the angle-resolution limitation of the telescope we used, it could not obtain the true peak value of its column density. It is possible that optical discs have higher column-density values in dwarf galaxies. We obtained that the ratio of gas to optical effective radius for KK153 is about 8.9, implying that the column density in the nearly entire structure ($\sim$99\%) of KK153 is lower than the aforementioned threshold value, indicating that the cosmic reionization background UV emission seems to have a limited impact on low-density \HI gases. The detection of KK153 poses a challenge to the theory of reionization.

\acknowledgments 
%We thank the referee for insightful comments that improved the clarity of this manuscript. 
We acknowledge the support of the National Key R$\&$D Program of China No. 2022YFA1602901. This work is also supported by the National Natural Science Foundation of China (Grant Nos. 12373001, 12225303, 12421003), the Chinese Academy of Sciences Project for Young Scientists in Basic Research, grant no. YSBR-063, the Guizhou Provincial Science and Technology Projects (QKHJC-ZK[2025]MS015), the Youth Innovation Promotion Association of CAS, and the Central Government Funds for Local Scientific and Technological Development (No. XZ202201YD0020C). 
This work made use of the data from FAST (https://cstr.cn/31116.02.FAST). FAST is a Chinese national mega-science facility, operated by the National Astronomical Observatories, Chinese Academy of Sciences.


\begin{thebibliography}{}

\bibitem[{{Adams} \& {Oosterloo}(2018)}]{Adams2018}
{Adams}, E. A. K., \&  {Oosterloo}, T. A., 2018, A\&A,  612, A26

\bibitem[{{Begum}  {et~al.}(2006)}]{Begum2006}
{Begum}, A., {Chengalur}, J. M., {Karachentsev}, I. D., et al. 2006, MNRAS, 365, 1220.

\bibitem[{{Ben{\'i}tez-Llambay}  {et~al.}(2017)}]{Benitez2017}
{Ben{\'i}tez-Llambay}, A., {Navarro}, J. F., {Frenk}, C. S., et al. 2017, MNRAS, 465, 3913.

\bibitem[{{Blitz} \& {Robishaw}(2000)}]{Blitz2000}
{Blitz}, L., \&  { Robishaw}, T., 2000, \apj,  541, 675

\bibitem[{{Bullock} \& {Boylan-Kolchin}(2017)}]{Bullock2017}
{Bullock}, J. S., \&  {Boylan-Kolchin}, M., 2017, ARA\&A,  55, 343

\bibitem[{{Davies}  {et~al.}(2006)}]{Davies2006}
{Davies}, J. I., {Disney}, M. J., {Minchin}, R. F., et al. 2006, MNRAS, 368, 1479.

\bibitem[{{Gallart}  {et~al.}(2021)}]{Gallart2021}
{Gallart}, C., {Monelli}, M., {Ruiz-Lara}, T., et al. 2021, \apj, 909, 192.

\bibitem[{{Giovanelli}  {et~al.}(2013)}]{Giovanelli2013}
{Giovanelli}, R., {Haynes}, M. P., {Adams}, E. A. K., et al. 2013, \aj, 146, 15.

\bibitem[{{Hunter} \& {Elmegreen}(2006)}]{Hunter2006}
{Hunter}, D. A., \&  {Elmegreen}, B. G. 2006, \apjs,  162, 49

\bibitem[{{Irwin}  {et~al.}(2007)}]{Irwin2007}
{Irwin}, M. J., {Belokurov}, V., {Evans}, N. W., et al. 2007, \apj, 656, L13.

\bibitem[{{Herrmann}  {et~al.}(2016)}]{Herrmann2016}
{Herrmann}, K. A., {Hunter}, D. A., {Zhang}, H. X., \& {Elmegreen}, B. G. 2016, \aj, 152, 177.

\bibitem[Herzog et al. (2023)]{Herzog2023} 
{Herzog}, G.,  {Ben{\'i}tez-Llambay}, A., \&  {Fumagalli}, M. 2023, MNRAS, 518, 6305


\bibitem[{{Jiang} {et~al.}(2019)}]{Jiang2019}
{Jiang}, P., {Yue}, Y. L., {Gan}, H. Q., et al. 2019, Sci. China-Phys.Mech. Astron. 62, 959502

\bibitem[{{Jiang} {et~al.}(2020)}]{Jiang2020}
{Jiang}, P., {Tang}, N.-Y., {Hou}, L.-G., et al. 2020, Research in Astronomy and Astrophysics, 20, 064

\bibitem[{{Jones} {et~al.}(2024)}]{Jones2024}
{Jones}, M. G., {Sand}, D. J., {Mutlu-Pakdil}, B., et al. 2024, \apjl, 971, L37.

%\bibitem[{{J{\'o}zsa}  {et~al.}(2017)}]{Jozsa2007}
%{J{\'o}zsa}, G. I. G., {Kenn}, F., {Klein}, U., {Oosterloo}, T. A., 2007, \aap, 468, 731.

\bibitem[{{Kauffmann}  {et~al.}(1993)}]{Kauffmann1993}
{Kauffmann}, G., {White}, S. D. M., \& {Guiderdoni}, B. 1993, MNRAS, 264, 201.

\bibitem[{{Kourkchi}  {et~al.}(2020)}]{Kourkchi2020}
{Kourkchi}, E., {Courtois}, H. M., {Graziani}, R., et al. 2020, \aj, 159, 67.

\bibitem[{{Klypin}  {et~al.}(1999)}]{Klypin1999}
{Klypin}, A., {Kravtsov}, A. V., {Valenzuela}, O., \& {Prada}, F. 1999, \apj, 522, 82.

\bibitem[{{Lelli}  {et~al.}(2016)}]{Lelli2016}
{Lelli}, F., {McGaugh}, S. S., \& {Schombert}, J. M., 2016, \apjl, 816, L14.

\bibitem[{{Lelli} (2022)}]{Lelli2022}
{Lelli}, F. 2022, Nature Astronomy, 6, 35

\bibitem[{{Mainzer}  {et~al.}(2014)}]{Mainzer2014}
{Mainzer}, A., {Bauer}, J., {Cutrl}, R. M., et al. 2014, \apj, 792, 30.

\bibitem[{{Moore}  {et~al.}(1999)}]{Moore1999}
{Moore}, B., {Ghigna}, S., {Governato}, F., et al. 1999, \apj, 524, L19.

\bibitem[{{McGaugh} (2012)}]{McGaugh2012}
{McGaugh}, S. S., 2012, \aj, 143, 40.

\bibitem[{{McGaugh} \& {Schombert}(2015)}]{McGaugh2015}
{McGaugh}, S. S., \&  {Schombert}, J. M., 2015, \apj,  802, 18

\bibitem[Neff et al. (2005)]{Neff2005} 
{Neff}, S. G.,  {Thilker}, D. A.,  {Seibert}, M., et al.  2005, \apj, 619, L91

\bibitem[Nelson et al. (2018)]{Nelson2018} 
Nelson, D., Pillepich, A., Springel,V., Weinberger, R., Hernquist, L. et al.  2018, MNRAS, 475, 624

\bibitem[Okamoto et al. (2008)]{Okamoto2008} 
{Okamoto}, T.,  {Gao}, L., \&  {Theuns}, T. 2008, MNRAS, 390, 920

\bibitem[{{Pan}  {et~al.}(2015)}]{Pan2015}
{Pan}, Z., {Li}, J., {Lin}, W., et al. 2015, \apjl, 804, L42.

%\bibitem[{{Ponomareva}  {et~al.}(2017)}]{Ponomareva2017}
%{Ponomareva}, A. A., {Verheijen}, M. A.~W., {Peletier}, R. F., {Bosma}, A., 2017, MNRAS, 469, 2387.

\bibitem[{{P{\'e}rez}  {et~al.}(2013)}]{Perez2013}
{P{\'e}rez}, E., {Cid Fernandes}, R., {Gonz{\'a}lez Delgado}, R. M., et al. 2013, \apjl, 764, L1.

\bibitem[Quinn et al. (1996)]{Quinn1996} 
{Quinn}, T.,  {Katz}, N., \&  {Efstathiou}, G. 1996, MNRAS, 278, L49

\bibitem[{{Ricotti} \& {Gnedin}(2005)}]{Ricotti2005}
{Ricotti}, M., \&  {Gnedin}, N. Y. 2005, \apj,  629, 259

\bibitem[{{Sales}  {et~al.}(2022)}]{Sales2022}
{Sales}, L. V., {Wetzel}, A., \& {Fattahi}, A. 2022, Nature Astronomy, 6, 897.

\bibitem[{{Simon} (2019)}]{Simon2019}
{Simon}, J. D. 2019, ARA\&A,  57, 375

\bibitem[{{Taylor}  {et~al.}(2017)}]{Taylor2017}
{Taylor}, R. {Davies}, J. I., {J{\'a}chym}, P., et al. 2017, MNRAS, 467, 3648.

\bibitem[{{Tollerud} {et~al.}(2008)}]{Tollerud2008}
{Tollerud}, E. J., {Bullock}, J. S., {Strigari}, L. E., \& {Willman}, B. 2008, \apj, 688, 277

\bibitem[{{Tollerud} {et~al.}(2015)}]{Tollerud2015}
{Tollerud}, E. J., {Geha}, M. C., {Grcevich}, J., {Putman}, M. E., \& {Stern}, D. 2015, \apjl, 798, L21

%\bibitem[{{Warren} {et~al.}(2012)}]{Warren2012}
%{Warren}, S. R., {Jurek}, R., {Obreschkow}, D. et al. 2014, MNRAS, 438, 1176

\bibitem[{{Wang} {et~al.}(2017)}]{Wang2017}
{Wang}, E., {Kong}, X., {Wang}, H. et al. 2017, \apj, 844, 144

\bibitem[{{Warren} {et~al.}(2012)}]{Warren2012}
{Warren}, S. R., {Skillman}, E. D., {Stilp}, A. M., et al. 2012, \apj, 757, 84

\bibitem[{{Westmeier} {et~al.}(2014)}]{Westmeier2014}
{Westmeier}, T., {Jurek}, R., {Obreschkow}, D., et al. 2014, MNRAS, 438, 1176

\bibitem[{{Xu} {et~al.}(2021)}]{Xu2021}
{Xu}, J. L., {Zhang}, C. P., {Yu}, N. et al. 2021, \apj, 922, 53

\bibitem[{{Young} \& {Lo}(1996)}]{Young1996}
{Young}, L. M., \&  {Lo}, K. Y. 1996, \apj,  462, 203

\bibitem[{{Young} \& {Lo}(1997)}]{Young1997}
{Young}, L. M., \&  {Lo}, K. Y. 1997, \apj,  490, 710

\bibitem[{Zhang} {et al.}(2023)]{Zhang2024}
{Zhang}, C. P., {Zhu}, M., {Jiang}, P., et al. 2024, Science China Physics, Mechanics,and Astronomy, 67, 219511

\end{thebibliography}
\end{document}